# Experimental realization of dice-lattice flat band at the Fermi level in layered electride YCl

Songyuan Geng[1], Xin Wang[1], Risi Guo[1], Chen Qiu[2], Fangjie Chen[1], Qun Wang[1], Kangjie Li[1], Peipei Hao[3], Hanpu Liang[2], Yang Huang[1], Yunbo Wu[4], Shengtao Cui[4], Zhe Sun[4], Timur K. Kim[5], Cephise Cacho[5], Daniel S. Dessau[3], Benjamin T. Zhou[1]*, Haoxiang Li[1]*

[1]Advanced Materials Thrust, The Hong Kong University of Science and Technology (Guangzhou), Guangzhou, 511453, China
[2]Department of Physics, Eastern Institute of Technology, Ningbo, 315200, China
[3]Department of Physics, University of Colorado, Boulder, CO, 80309, USA
[4]National Synchrotron Radiation Laboratory, University of Science and Technology of China, Hefei, Anhui 230029, China
[5]Diamond Light Source, Harwell Science and Innovation Campus, Didcot OX11 0DE, UK
*Corresponding author. Email: tongz@hkust-gz.edu.cn and haoxiangli@hkust-gz.edu.cn

**Abstract: Flat electronic bands, where interactions among electrons overwhelm their kinetic energies, hold the promise for exotic correlation physics. The dice lattice has long been theorized as a host of flat bands with intriguing band topology. However, to date, no material has ever been found to host the characteristic flat bands of a dice lattice. Here, using angle-resolved photoemission spectroscopy (ARPES), we discover a dice-lattice flat band at $E_F$ in the van der Waals (vdW) electride $[YCl]^{2+}: 2e^-$. In this system, excess valence electrons from Y deconfine from the cation framework to form an interstitial anionic electron lattice that constitutes the dice lattice. Our ARPES measurements unambiguously identify two sets of dice-lattice bands in YCl, including a nearly dispersionless band at the Fermi level. The flat bands and other dispersive bands observed in ARPES find excellent agreement with first-principles calculations, and theoretical analysis reveals that the near-$E_F$ electronic structure is well captured by a simple dice-lattice model. Our findings thus end the long quest of a real dice flat band material and establish vdW electride YCl as a prototype of dice metals. Our results further demonstrate the anionic electron lattice as a novel scheme for realizing lattice geometries and electronic structures rare to find in conventional crystalline systems.**

**Introduction**

Flat bands—non-dispersive electronic states with quenched kinetic energy—have emerged as a central concept in quantum materials research [1-7]. When positioned at the Fermi level ($E_F$), flat bands exhibit a spiked density of states that enhances electron interactions, driving diverse correlated quantum phases, such as superconductivity [8, 9], magnetism [10], and fractional quantum Hall states [11, 12]. Engineering lattice geometries represents a key strategy for realizing flat bands. Over the past decade, most experimental efforts have focused on the kagome and Moiré lattices, which have been successfully realized in several material systems [13-20].

Among lattices that host flat bands, the dice lattice stands out as a prototypical construct that has remained elusive in real materials since its proposal by Sutherland in 1986 [21]. Geometrically, the dice lattice consists of two types of lattice sites: the six-fold coordinated C sites (red circles in Fig. 1a) and the three-fold coordinated A/B sites (blue circles in Fig. 1a). In an ideal dice lattice, the C sites equivalently couple with all adjacent A/B sites, whereas the direct couplings between the neighboring A and B sites are broken [21-23]. In the electronic states of the dice lattice, one of the eigenfunctions is fully localized on the A/B sites and exhibit zero dispersion, as shown in Fig 1b, with two dispersive bands that intersect the flat band and form a Dirac cone. Owing to the flat band framework, the dice lattice has attracted considerable theoretical interest, with predicted correlated phases including flat band ferromagnetism [24], fractional Chern insulators [25], chiral superconductivity [26], interaction-driven quantum anomalous Hall effect [27], Aharonov-Bohm caging-induced insulators [28] and so on.

Despite theoretical promise, no material realizes the dice-lattice bands due to its strict geometric and energetic constraints, which requires all lattice sites (A, B and C sites) to host electronic orbitals within a proximate energy window [25, 29]. For instance, 1T-phase transition metal dichalcogenides adopt a dice-like lattice geometry; however, their alternating anion/cation sublattice occupation leads to large on-site energy differences [30, 31], hindering the formation of dice-lattice bands. A true ionic dice lattice demands same-charge ions on all sublattice sites to minimize energy mismatch, but the inherent electrostatic repulsion renders such structures unstable and rare to find in nature. Heterostructures and artificial lattices have been proposed[32-35], yet to date, no successful experimental attempt has been made along this line.

A promising route to overcome these limitations lies in the emerging class of two-dimensional (2D) electrides [36-38]. These materials, such as $[Ca_2N]^+:e^-$ [36, 39], $[Y_2C]^{2+}:2e^-$ [40,

41] and [YCl]$^{2+}$ : 2e$^-$ [42, 43] , are characterized by excess electrons that act as anions and are spatially confined within interlayer gaps or channels between positively charged cationic frameworks. These interstitial anionic electrons (IAEs) behave as quasi-metallic species, which circumvent the electrostatic penalties associated with densely packed metallic ions [44-48]. This electrostatically favorable arrangement provides a promising platform for realizing complex electron lattices [49-51], which enables the engineering of lattice geometries that are unlikely to achieve in conventional ionic lattices [40, 52, 53]. Further, being largely detached from the host ions, these weakly bound IAEs typically give rise to low-energy electronic states near $E_F$ [36, 39, 54]. Thus, when combined with frustrated lattice geometry, these anionic electron lattices offer a viable route to achieving exotic flat band structure near $E_F$. The van der Waals (vdW) 2D electride [YCl]$^{2+}$ : 2e$^-$ exemplifies such an approach [42, 43, 55]. In YCl, the anionic electrons are confined within interstitial sites that arrange into an electron lattice, which opens up the possibility of forming a dice lattice structure (Fig. 1c-e).

Here, we report the realization of a dice-lattice flat band at $E_F$ in the vdW 2D electride YCl using angle-resolved photoemission spectroscopy (ARPES). Our ARPES measurements resolve two sets of dice-lattice bands: one featuring a flat band at $E_F$ and another one with the flat band located approximately 600 meV below $E_F$. The near-$E_F$ electronic structure of YCl, comprising these flat bands and two dispersive band features observed by ARPES, agrees well with the density functional theory (DFT) prediction and mirrors the characteristic band structure of a dice lattice model. In YCl, the electronic states near $E_F$ are dominated by the IAEs [42, 43] (Fig. 1g). These IAEs located at the A, B and C sites form a unique lattice architecture (Fig. 1c-e). Crucially, direct hopping between the A and B-site IAEs is strongly suppressed, providing the essential geometric constraint that enables the formation of dice-lattice flat bands (Fig. 1a-e). Comparing ARPES results with theoretical calculations, our work demonstrates that the anionic electron lattice (Fig. 1c-e) in YCl gives rise to the dice-lattice electronic structure near $E_F$, establishing YCl as a dice metal. This material serves as a prototype of a broader class of rare-earth metal halide electrides ReX (Re = rare-earth metal and X = halogen), in which similar exotic flat band structures are theoretically anticipated. Our findings further demonstrate the power of anionic electron engineering in designing exotic lattice geometries and electronic structures, opening new avenues in the search for emergent quantum phenomena.

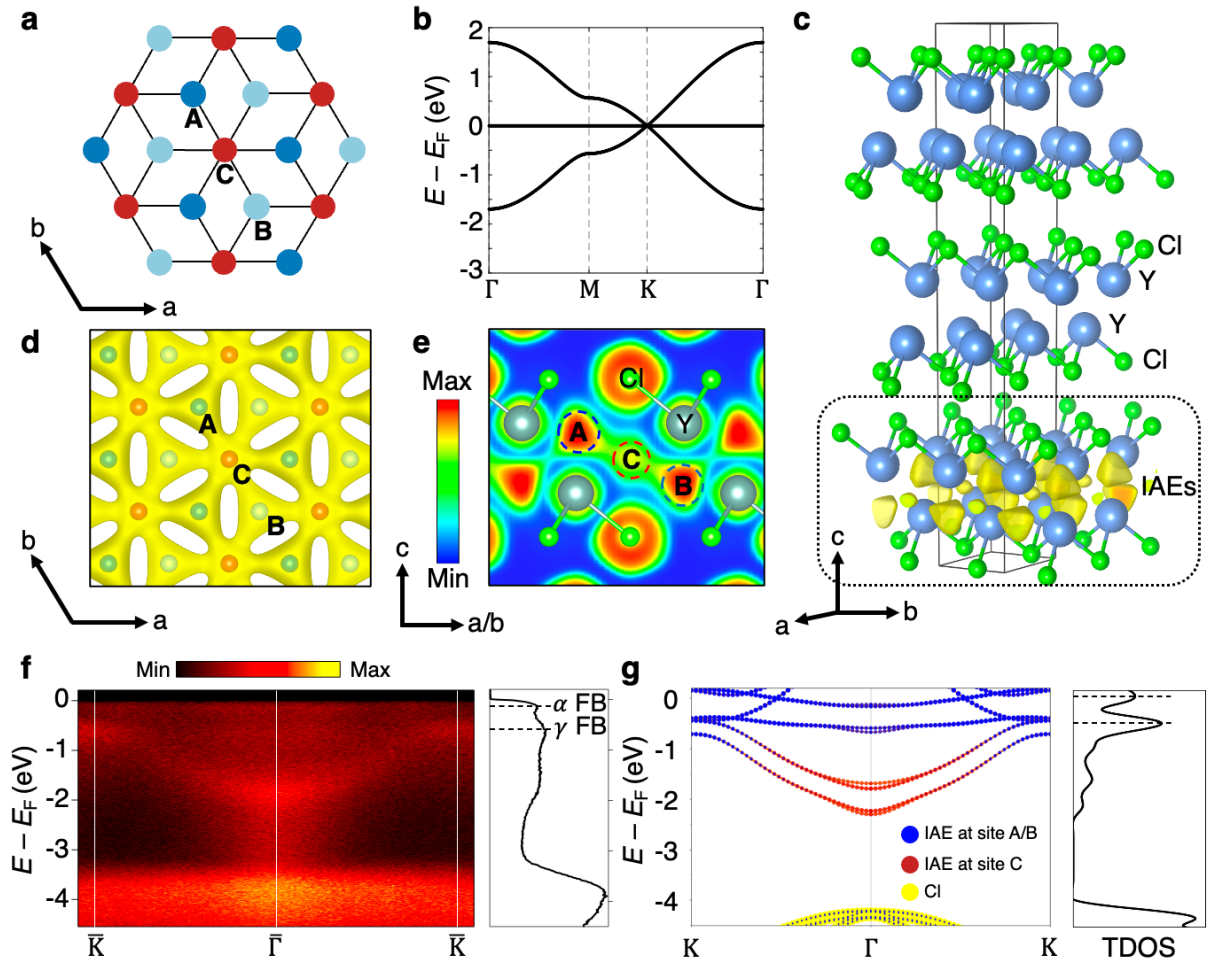

**Figure 1 Characteristics of dice lattice and 2D vdW electride YCl. a.** Schematic of dice lattice structure, composed of two types of lattice sites: A and B sites (dark blue and light blue) are three-fold coordinated with C sites, and each C site is six-fold coordinated with all neighboring A/B sites. **b.** Calculated band structure of an ideal dice lattice, showing a perfect flat band intersecting two dispersive bands at the K point. **c.** Crystal structure of YCl. The conventional unit cell consists of three alternating layers of Cl-Y-Y-Cl slabs, stacked along the *c*-axis, with the dashed box highlighting the monolayer. At the bottom layer of Cl-Y-Y-Cl slab, the interstitial anionic electrons (IAEs) confined between the Y layers are revealed by the electron localization function (ELF, isosurface = 0.6). **d.** Top-down view of the ELF (isosurface = 0.6), showing IAEs at sites A and B that are three-fold coordinated with adjacent C sites (no direct A-B bond), while site-C IAEs are six-fold coordinated with neighboring A/B sites. The ELF center-of-maxima are traced and marked with corresponding color to panel a. **e.** Side view of ELF (isosurface = 0), displaying the spatial distribution of IAEs at A, B and C sites. **f.** Angle resolved photoemission spectroscopy (ARPES)

spectrum and the integrated spectral weight of the ARPES cut, revealing two flat bands near $E_F$. The two peaks near $E_F$ in the integrated spectral weight point out the positions of these two flat bands ($\alpha$ and $\gamma$ flat bands). **g.** DFT band structure and the total density of states of YCl. The near-$E_F$ states are dominated by IAEs at A/B sites (blue dots) that originate from Y $4d_{z^2}$ electrons and IAEs at C sites (red dots) that originate from Y $5s$ electrons.

**Dice lattice formed by the interstitial anionic electrons in YCl**

YCl adopts a rhombohedral phase (space group $R\bar{3}m$, point group $D_{3d}$) at ambient conditions (Fig. 1c), featuring a layered Cl–Y–Y–Cl slab structure stacked along the $c$-axis (see crystal structure information in supplementary materials S1). This arrangement turns YCl into a vdW material, in which the interlayer Cls are held together by a weak vdW force. In YCl, Y donates one electron to Cl and transfers its remaining two valence electrons to the interstitial voids [42, 43] (see differential charge density diagram in supplementary materials Fig. S4), making the overall oxidation state of $Y^{3+}$. X-ray photoemission spectroscopy (XPS) measurement confirms the trivalent oxidation state of Y in our sample (see supplementary materials Fig. S5), which is consistent with a previous study on YCl [43]. These excess electrons in YCl occupy two spatially distinctive sites (site A/B and site C), as shown by the electron localization function (ELF) maps in Fig. 1d,e. The IAEs located at sites A and B originate from Y $4d_{z^2}$ electrons, while the IAEs at site C located at the center of each hexagon originate from the Y $5s$ electrons (Fig. 1g, and more details in supplementary materials S2). This arrangement of IAEs in YCl gives rise to an electron lattice framework that mirrors the architecture of the dice lattice (Fig. 1a,d).

With the IAEs forming a dice lattice, electronic states arising from the IAEs present characteristics of the dice lattice bands. In particular, two nearly dispersionless bands are identified near the Fermi level in the ARPES spectrum (Fig. 1f), consistent with the DFT calculation (Fig. 1g). These two bands of narrow bandwidths corresponds to the peaks in the integrated spectral weight near $E_F$ in ARPES and the pronounced peaks in the DFT-derived density of states. Since trivalent Y has a closed-shell configuration, the IAEs dominate the near-$E_F$ electronic structure, as confirmed by the DFT fat-band plot. In addition, the near-$E_F$ IAEs states are well-isolated in energy, where the localized Cl bands are separated by a substantial energy gap of over 1.5 eV, which minimizes hybridization and suppresses coupling between the Cl states and the near-$E_F$ bands.

## ARPES observation of dice-lattice bands

The near-$E_F$ electronic structure of YCl is highlighted in the ARPES spectra as shown in Fig. 2a and further revealed by the dispersions extracted from the peak positions of the energy distribution curves (EDCs) in Fig. 2e. In particular, the two flat bands near $E_F$ are consistent with the narrow bands originating from the IAEs at sites A/B, as evidenced by the band-projected electron wavefunction (Fig. 2b,c). In contrast, the strongly dispersive band at lower energy (with the band minimum at –1.7 eV) arises from the IAEs at site C (Fig. 2b,d). These key band features below $E_F$, which are observed in both ARPES and DFT results, can be described by a dice-lattice tight-binding (TB) model (Fig. 2f-h), constructed based on the IAE geometry derived from the ELF map (Fig. 1d).

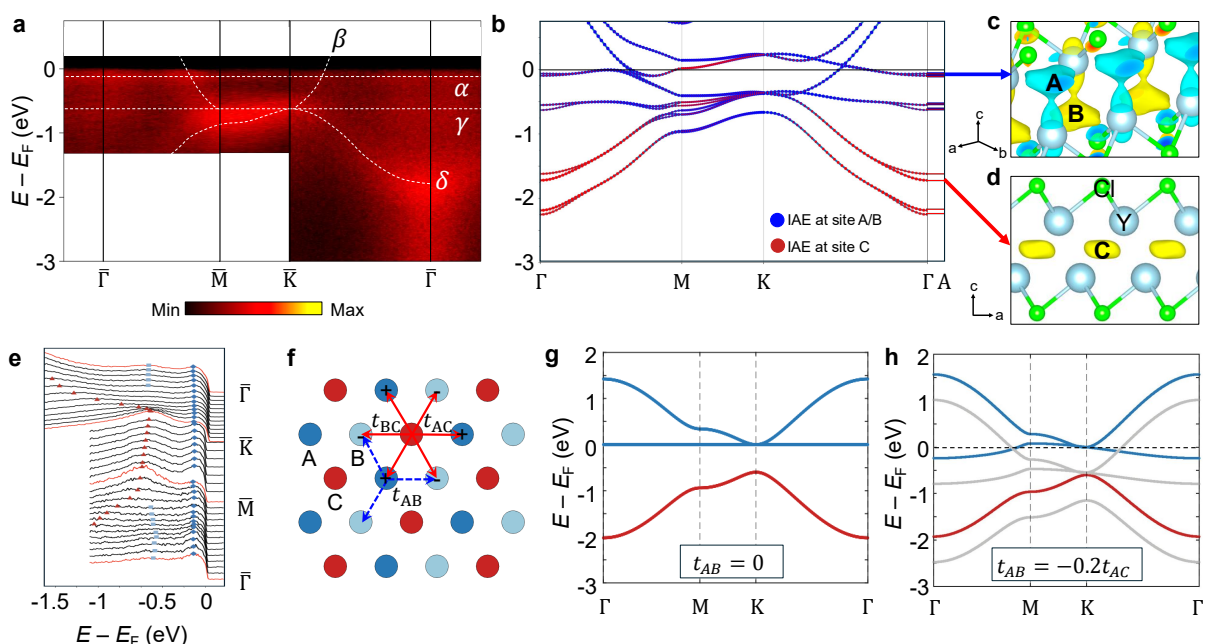

**Figure 2 Experimental and computational electronic structure of dice lattice. a.** ARPES spectra of YCl along the high-symmetry directions. The white dashed lines are the guide-to-the-eye lines that highlight the flat bands ($\alpha$ and $\gamma$) and the dispersive bands ($\beta$ and $\delta$), which form two sets of the dice-lattice bands below $E_F$. The ARPES data is taken with 75 eV photons at temperature $T$ = 6 K. **b.** DFT band structure of YCl bulk crystal, with the IAE states at sites C indicated by blue dots, and the IAE states at sites A/B indicated by red dots. **c.** The band-projected wavefunction of the $\alpha$ flat band. The yellow and blue isosurface represent the positive and negative sign of the wavefunction lobes, respectively. **d.** The band-projected wavefunction of the dispersive $\delta$ band. The wavefunction associated with the dispersive $\delta$ band is completely isolated from the ionic cores

and locates at the inversion center of each monolayer Cl-Y-Y-Cl slab. **e.** Stacking plot of energy dispersion curves (EDCs) along the high symmetry cuts as shown in panel **a**. The dispersions extracted from the EDC peak positions reveal the $\alpha$ flat band (dark blue diamond), the $\gamma$ flat band (light blue square) and the dispersive $\delta$ band (red triangle). **f.** Geometry of the dice-lattice tight-binding (TB) model, illustrating the dominant hopping terms $t_{AC}$, $t_{BC}$ (solid red arrows). The suppressed hoppings between A and B sites ($t_{AB}$) are indicated by the dashed arrows. The dice lattice bands are generated with the TB model using $t_{AB} = 0$ (panel **g**) and $t_{AB} = -0.2 t_{AC}$ (panel **h**). The grey bands in panel h represent duplicate dice bands with an energy offset of -0.6 eV, included for comparison with the two sets of dice bands from the DFT calculation.

In an ideal dice lattice, the on-site energies of A, B and C sites are identical, with neighboring bonds between A and B sites being broken [21-23]. The AB-sublattice antibonding states of odd parity are fully decoupled from the parity-even states on site C, regardless of the hybridization between electrons at the A/B site and the central C site. Combining with the broken bond between the A and B site, the kinetic energy of the AB-sublattice antibonding states are thus quenched, leading to a dispersionless band and a three-fold band crossing at the K point as shown in Fig 1b [21, 25].

The IAE architecture of YCl deviates slightly from an ideal dice lattice, where the on-site energy of site-C electrons is approximately 0.6 eV lower than that of site-A and B electrons. The on-site energy difference gives a TB band structure (Fig. 2g) that is comparable to the ARPES and DFT results (Fig. 2c), and reduces the three-fold band crossing at the K point to two-fold [27, 56] (Fig. 1b, Fig. 2g). Nevertheless, the dice flat band remains intact under the sublattice asymmetry between site C and sites A/B (Fig. 2g). This is because the on-site energy difference between A/B and C does not break spatial inversion symmetry (with site C defined as the inversion center).

In addition, the IAEs of YCl at site A and B exhibit electronic wavefunctions resembling the dumb-bell shape of $d_{z^2}$ orbital (Fig. 2c). The alternating signs of the wavefunction lobes restrict the direct hopping between the A and B sites, providing the essential suppression of the A-B bonds. While a perfect dice lattice requires fully broken A-B bonds, realistically the hopping between the A and B sites may not be fully quenched. A finite but small hopping parameter $t_{AB}$ can generate weakly dispersed band features, which can account for the small but finite bandwidth shown in the DFT result (Fig. 2h). Nevertheless, the ARPES data present an even flatter dispersion compared to the DFT bands, revealing that the YCl electride is well in the dice-lattice limit with nearly broken bonds between A and B sites.

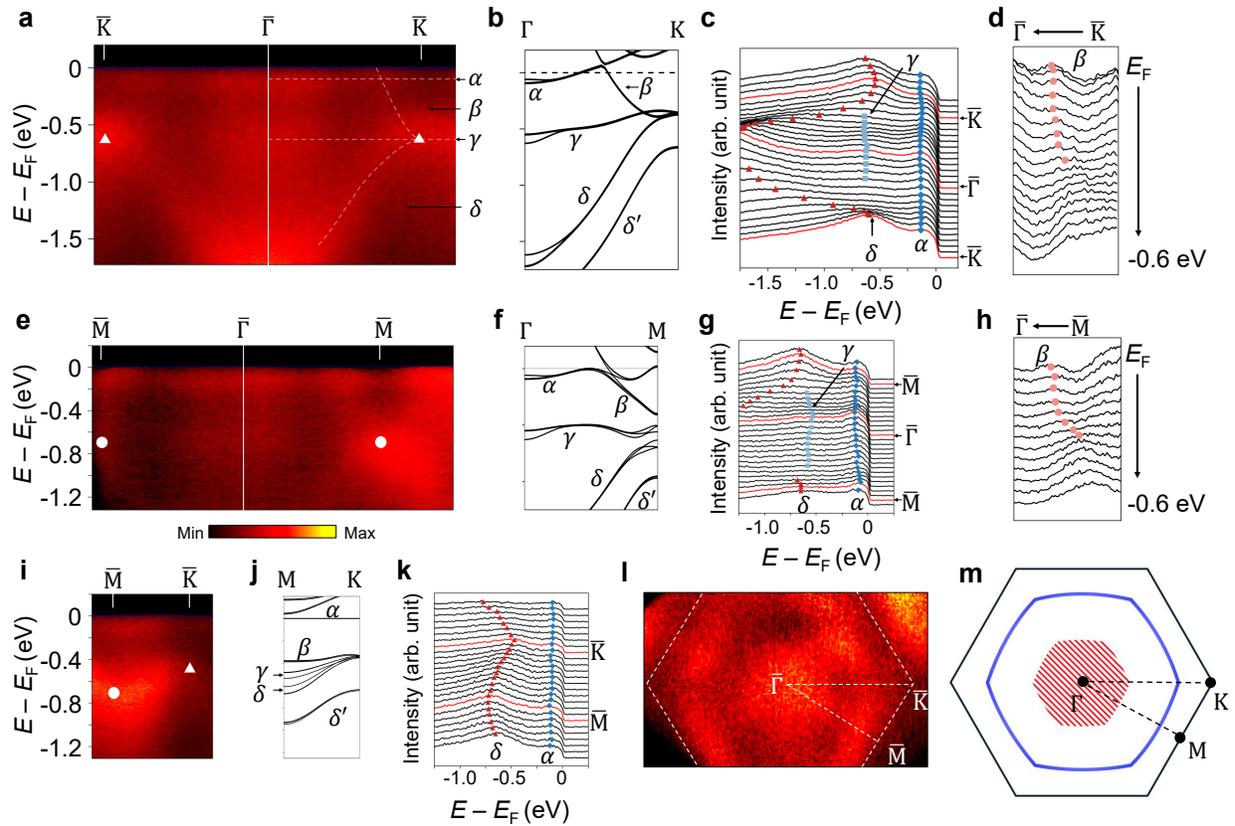

**Figure 3 Detailed analysis of the electronic structure of YCl. a.** ARPES spectrum measured along the high-symmetry $\overline{K} - \overline{\Gamma} - \overline{K}$ direction, revealing the four bands near $E_F$: the top flat band (α band), the lower flat band (γ) and two dispersive bands (β and δ). The white triangle marks the band top of the δ band at $\overline{K}$. **b.** DFT-calculated band structure along the same momentum path of panel a, exhibiting excellent agreement with the experimentally observed band features. **c.** Stacking EDCs along the $\overline{K} - \overline{\Gamma} - \overline{K}$ path. The dispersions of α (dark blue diamond), γ (light blue square) and δ (red triangle) bands are extracted from the EDC peaks. **d.** Stacking momentum distribution curves (MDCs) of panel a, with the β band dispersion (pink circle) extracted from the MDC peaks. **e-h.** ARPES spectrum, DFT bands, stacking EDCs and MDCs plots along the $\overline{M} - \overline{\Gamma} - \overline{M}$ direction, illustrating consistent band features and dispersion behavior. **i-k.** ARPES spectrum, DFT bands and stacking EDCs plot along $\overline{M} - \overline{K}$ direction. **l.** Fermi surface map of YCl. **m.** Schematics of the Fermi surface, showing the hole pocket formed by the β band (blue line) and the enhanced spectral weight from the $\alpha$ band around the $\bar{\Gamma}$ point as revealed by the ARPES result in **l**. All the ARPES data were taken with 75 eV photons at sample temperature $T = 6$ K.

The ARPES spectra and Fermi surface map in Fig. 3 provide a comprehensive view of the electronic structure. Along high-symmetry cuts (Fig. 3a-k), the four bands arising from the IAEs ($\alpha$, $\beta$, $\gamma$ and $\delta$) are observed. EDC stacking plots (Fig. 3c,g,k) highlight the dispersionless nature of the $\alpha$ and $\gamma$ bands, contrasting with the dispersive $\delta$ band. The $\beta$ band, which cannot be identified from the EDCs due to the highly dispersive nature with large band velocity, is evident in the stacking plot of the momentum distribution curves (MDCs) (Fig. 3d,h). Notably, the $\beta$, $\gamma$ and $\delta$ bands exhibit an accidental degeneracy near the zone corner $\bar{K}$ point at around -0.55 eV (triangle marks in Fig. 3a,i). Along the $\bar{M}$-$\bar{K}$ cut (Fig. 3j), DFT predicts the $\alpha$ band residing above the Fermi level, while the ARPES reveals substantial spectral weight below $E_F$, with spectral peaks resolved in the stacking EDC plot (Fig. 3k). This is further supported by the $\bar{\Gamma}$-$\bar{K}$ and $\bar{\Gamma}$-$\bar{M}$ cuts (Fig. 3c,g), where spectral peaks right below $E_F$ appear near both $\bar{M}$ and $\bar{K}$ points. The ARPES data conclusively show that the $\alpha$ band is even flatter than the DFT prediction, forming a truly dispersionless feature.

The Fermi surface map shown in Fig. 3l reflects the strong spectral contributions from the nearly flat $\alpha$ band at the Fermi level, resulting in broad and intense spectral features. Two prominent regions of spectral weight can be resolved. One of these features centers around the $\bar{\Gamma}$ point, arising from the enhanced spectral intensity of the dispersionless $\alpha$ band near the $\bar{\Gamma}$ point, consistent with the ARPES results shown in Fig. 3a,e. The other feature exhibits spectral intensity around the zone boundary (Fig. 3l,m), arising from the intersection of the dispersive $\beta$ band and the $\alpha$ flat band at the $E_F$ (Fig. 3a,e), where the $\beta$ band forms a large hole pocket centered around the $\bar{\Gamma}$ point as indicated in the schematic Fermi surface plot in Fig. 3m. Both features on the Fermi surface present intensity modulation of the hexagon shape (Fig. 3l), consistent with the symmetry of the YCl crystal. Fermi surface from DFT, on the other hand, presents complicated structure due to the wiggling of the $\alpha$ flat band around the Fermi level (see supplementary materials Fig. S8).

To further examine the flat band structure of YCl, we present ARPES spectra that cut across different Brillouin zone (BZ) regions in Fig. 4a-c. Two sets of ARPES cuts—parallel cuts to $\bar{\Gamma}$-$\bar{M}$ direction and to $\bar{\Gamma}$-$\bar{K}$ direction (with the cut position indicated in Fig. 4a)—reveal the two flat band features that span the entire BZ. The α band and the γ band exhibit dispersionless features throughout the BZ, indicating strong quenching of the electron kinetic energy. In addition to the dispersionless bands, the parallel cuts shown in Fig. 4b,c present well distinguishable dispersions along the path, showing a continuous evolution of the dispersive dice bands across the BZ. Overall, the ARPES band features are consistent with the evolution of the dice lattice bands as shown by the curvature plot of the dice band structure in supplementary materials Fig. S12. As a vdW layered material, YCl is predicted to host a quasi-2-dimensional electronic structure with negligible $k_z$ dispersion (Fig. 2b). Although there are strong matrix element effects in the photon-energy dependent measurement of YCl (supplementary materials Fig. S10), our ARPES data displays consistent flat band and dispersive band features with different photon energies.

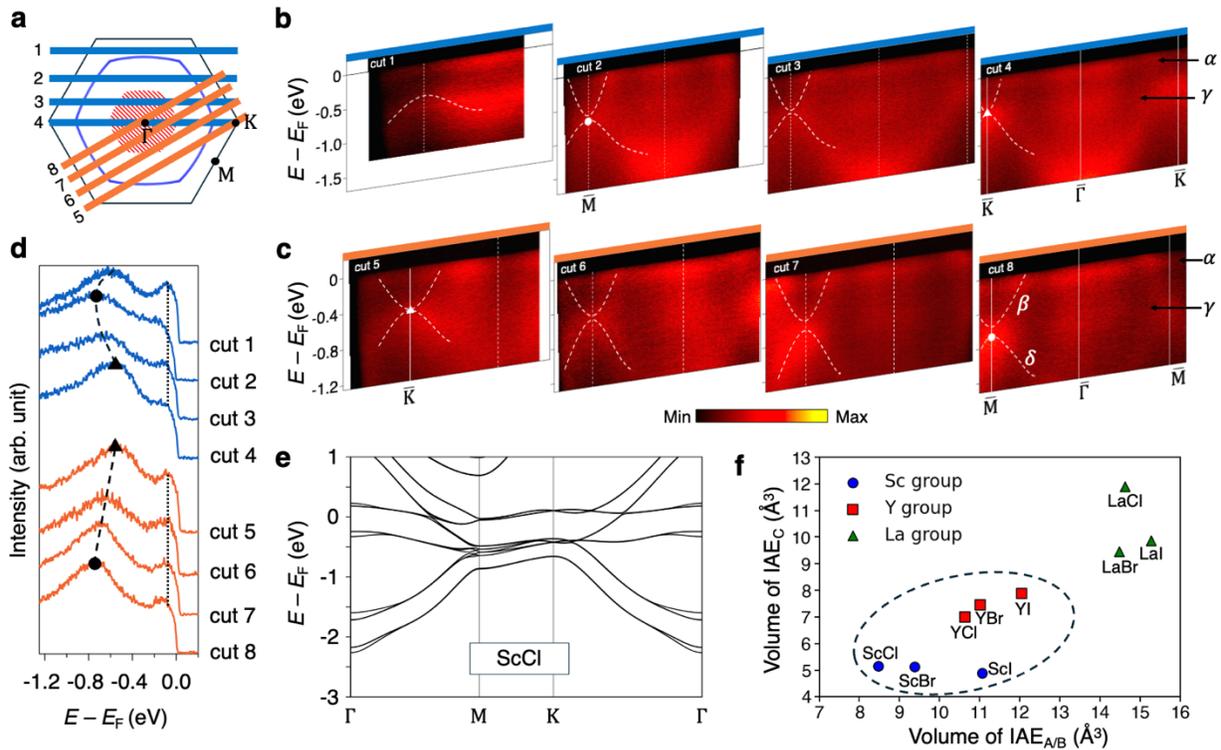

**Figure 4 ARPES cuts across the Brillouin zone and comparison with other ReX crystals. a-c.** Two sets of ARPES spectra: cuts 1-4 (blue) taken parallel to the $\overline{K} - \overline{\Gamma} - \overline{K}$ and cuts 5-8 (orange) taken parallel to the $\overline{M} - \overline{\Gamma} - \overline{M}$ direction. The cut positions are indicated in panel a. The α and γ flat bands near the Fermi level are clearly resolved in all cuts, accompanied by the dispersive β and δ bands. The white dashed lines serve as the guide-to-the-eye lines, indicating the band dispersions. **d.** EDCs from cuts 1-8 taken from the zone boundaries, showing the consistent peak

positions of the α flat band (black dotted line) and the evolution of dispersion peaks for δ band (black dashed line) with the triangle and circle mark the δ-band maximum and minimum along the zone boundary, respectively. **e**. DFT band structure of ScCl, showing two sets of dice flat bands similar to that of YCl. **f**. $IAE_{A/B}$ vs $IAE_C$ volumes for ReX electrides (Re = Sc, Y, La; X = Cl, Br, I). Blue circles, red squares, and green triangles denote Sc-, Y-, and La-based compounds, respectively. Electrides in the small-volume region (dashed ellipse), the Sc- and Y-based electrides, exhibit dice-lattice flat bands similar to those of YCl, whereas the La-based compounds do not. Band structures are provided in supplementary materials Fig. S13.

The ARPES spectra of YCl in Fig. 1–4 exhibit large broadening across all bands. This spectral broadening is potentially induced by the two dispersionless bands. With the diminishing band velocity, even modest electron self-energy could introduce a significant broadening effect in the spectral function [57]. As all the near-$E_F$ bands in YCl arise from the IAE states, the presence of the electronic screening effect [58] from the IAEs could potentially provide substantial electron self-energy. The observed broadening could also stem from charge inhomogeneities. Previous study indicates that the IAE states are susceptible to the crystal environment [59]. Thus, defects and impurities could readily perturb the IAEs and cause spectral broadening on the IAE bands.

Unlike the two well-separated flat bands, the dispersive δ band displays a single prominent dispersion. We assign the higher-energy split component in DFT as the δ band, given it merging with γ flat band at the K point, consistent with ARPES observation. The lower-energy split component predicted by DFT (δ′ band), however, is absent in the ARPES spectra. To further investigate the absence of the δ band splitting, we explore the matrix element effect with different photon energies (see supplementary materials Fig. S10). Though there are quite strong modulations of the ARPES intensity with different photon energies, the single δ-band dispersion remains unchanged. It is worth noting that the splitting between the two sets of dice bands in DFT is largely due to the spin-splitting within the monolayer (see discussion of the spin structure in supplementary materials S3). The splitting of the flat band due to Stoner instability has been predicted by the Hubbard model, where the splitting of dispersive bands is less affected compared with that of the flat band [60]. In our case, DFT calculations predict that the splitting of the δ band is approximately 0.25 eV, about half that of the flat bands. In the real material, the δ band splitting could be even smaller than the DFT prediction and thus, not resolvable in the ARPES spectra.

Comparing the ARPES data to the DFT result and the TB model simulation shown in Fig. 2, YCl exhibits a remarkably simple electronic structure with two sets of dice-lattice bands near the Fermi level, free from complications or contamination by other electronic states. This is in stark contrast with most kagome intermetallic compounds [5, 61, 62], where multiple metallic ions contribute to the low-energy electron density. This energetic separation of the Y-derived valence states near $E_F$ makes YCl an ideal, simplified flat-band system for study.

Our analysis shows that the formation of dice flat bands in electrides requires two key ingredients: (i) IAEs at site C that hybridize with those at sites A and B, and (ii) sufficiently weak direct A–B hopping. Both conditions emerge naturally from the lattice architecture of IAEs in YCl. As a proof of principle, Fig. 4e–f presents a DFT survey of other ReX crystals that can also host IAEs. We find that dice-lattice flat bands are not limited to YCl but also appear in a wider family of Sc- and Y-based electrides, demonstrating that dice-band physics is robust across this class of materials. In contrast, the La-based electrides, with the spatially extended IAEs derived from the $5d$ electrons, fail to preserve the dice lattice bands (see the DFT band structures in supplementary materials S6). Overall, these results establish ReX electrides as a promising platform for realizing and engineering dice-lattice flat band materials.

**Discussion**

Traditionally, the lattice points of a crystal are defined by the positions of atoms, ions, or molecules. However, in YCl, we uncover a fundamentally distinct scenario: the anionic electrons themselves could occupy well-defined lattice sites, acting as emergent quantum constituents that establish a dice-lattice structure. This electron-centric lattice does not originate from a distortion or modulation of the underlying atomic scaffold [36, 38, 41, 54]; rather, it emerges naturally from the electrostatic environment provided by the Y and Cl ions [42]. The symmetry of the host lattice is fully preserved, and the translational invariance of the unit cell remains intact. This sharply contrasts with the charge density wave (CDW), where electron density modulation forms a superlattice that breaks the original symmetry [63-65]. In addition, strong electron interactions induce Wigner crystals that feature spatially ordered electrons arising from electron fluids independent of atomic lattices [66-69]. In contrast, the electride state in YCl is crystal-encoded—the voids of the YCl structure define discrete quantum wells that host anionic electrons without spontaneous symmetry breaking as in Wigner crystallization and CDW. Therefore, YCl represents a new paradigm where anionic electrons act as periodic building blocks, effectively becoming "electronic ions" embedded within a crystalline potential. This opens a conceptual bridge between electride physics and exotic lattice geometries, such as the dice lattice.

In summary, we report the experimental realization of a dice-lattice flat band at the Fermi level in the vdW electride YCl. Leveraging the intrinsic anionic electron lattice formed by excess Y valence electrons, we uncover a dice lattice geometry that gives rise to an idealized flat band electronic structure, long predicted but previously non-existent in crystalline materials. Our findings establish YCl as a prototype dice metal and highlight electrides as a fertile platform for engineering exotic lattice geometries through anionic electron design. This work not only ends the decades-long quest for a real-material host of dice flat bands but also opens a conceptual avenue for realizing exotic electronic structures in electron-configured lattices.

**Methods**

**Synthesis:** Single crystals of YCl were synthesized via a self-flux method using stoichiometric mixtures of yttrium (Y) and yttrium trichloride ($YCl_3$). The Y and $YCl_3$ powders were ground together using an agate mortar and pestle, then pressurized into pellets. Each pellet was encapsulated within a molybdenum capsule and then further sealed in a stainless-steel capsule. All these preparation works were conducted in a glove box. The assembly was heated at 1050°C for 1.5 days, then cooled down to 800 °C for 6 days under Ar flow.

**ARPES Measurements:** ARPES data shown in this paper were collected using beamline I05 at Diamond Light Source[70] with vacuum pressure better than $1.3 \times 10^{-10}$ mbar. All samples were cleaved in-situ to ensure clean surfaces. ARPES data shown in the main text were taken with photon energy $hv$=75 eV under linear-horizontal polarization, with the sample temperature maintained at $T$ = 6 K. Measurements were conducted using an MBS A-1 electron analyzer, where the total energy resolution is about 5 meV. All Fermi surface maps presented in this study were integrated over an energy range of $E_F \pm 10$ meV.

**DFT Calculations:** First-principles calculations based on Density Functional Theory (DFT) were performed using the Vienna Ab initio Simulation Package (VASP)[71, 72], employing the projector augmented-wave (PAW) formalism[73-75]. The local density approximation (LDA) functional, following the Ceperley-Alder parameterization as reformulated by Perdew and Zunger (CAPZ)[76, 77], was adopted for the exchange-correlation energy. A Monkhorst-Pack $k$-point grid was employed with a typical density of 12 × 12 × 3 for bulk calculations, and an energy cutoff of 600 eV was used for the plane-wave basis. Structural relaxations were performed for all magnetic configurations until the residual Hellmann-Feynman forces on each atom were less than 0.002 eV/Å, and total energy was converged to better than $10^{-6}$ eV. All calculations were carried out in a spin-polarized setting to account for possible magnetic orderings in bulk YCl. All IAEs' charge states and centers of maximum are calculated using BadELF with the zero-flux method for separating metallic features. The searching criteria for electrode charge is set to 0.6[78].

We investigated both ferromagnetic (FM) configurations using the conventional unit cell and type-A antiferromagnetic (AFM) configurations with a 1 × 1 × 2 supercell by setting layer-alternating initial magnetic moments on Y ions (see comparison in supplementary materials Fig. S7). The overall band structures of two magnetic configuration are largely consistent, but type-A AFM configuration provides band structure that aligns better to the ARPES data and results in lower ground state total energy compared to the FM configuration. Thus, in the main text, we compare the ARPES data to the DFT calculations with the AFM configuration. The calculation of all ReX electrides follows that of YCl, using the 1 × 1 × 2 supercell with initial magnetic moments.

**Three-band Tight Binding Model:** Details of the three-band TB Hamiltonian used to illustrate the flat band physics are presented below: in the Bloch basis of the A, B and C sublattice sites, the full $\bm{k}$-space Hamiltonian can be written as the following 3-by-3 matrix:

$$H(\bm{k}) = \begin{pmatrix} h_A(\bm{k}) & h_{AC}(\bm{k}) & h_{AB}(\bm{k}) \\ h_{AC}^*(\bm{k}) & h_B(\bm{k}) & h_{BC}(\bm{k}) \\ h_{AB}^*(\bm{k}) & h_{BC}^*(\bm{k}) & h_C(\bm{k}) \end{pmatrix}, \qquad \text{Eq. 1}$$

where $h_l(\bm{k}) = E_l$ with site index $l = A, B, C$ describes the on-site energy $E_l$ for site A, B and C, respectively; $h_{AC}(\bm{k}) = t_{AC} f_0(\bm{k}) = h_{BC}(\bm{k}) = t_{BC} f_0(\bm{k})$, with the geometric factor $f_0(\bm{k}) \equiv \sum_{j=1,2,3} e^{i\bm{k}\cdot\bm{\delta}_j}$, denote the hopping terms between site A/B to site C characterized by parameter $t_{AC}$ and $t_{BC}$, where $\bm{\delta}_j$ is the hopping vectors; $h_{AB}(\bm{k}) = t_{AB} f_0^*(\bm{k})$ denotes the direct coupling between site A and site B IAEs with hopping parameter $t_{AB}$. $H(\bm{k})$ satisfies all symmetry constraints imposed by the $D_{3d}$ point group symmetry of the electride system. More details on the spin-polarized TB model can be found in supplementary materials S6.

**Acknowledgements**

The work conducted at Hong Kong University of Science and Technology (Guangzhou) was supported by NSFC-Young Scientists Fund (No. 12304093, No. 12447158), Guangzhou Basic and Applied Basic Research Scheme (No. 2024A04J4509), and Start-up Fund of HKUST(GZ) through Grant No. G0101000127 and No. G0101000263. The Modern Matter Laboratory (MML) and the Materials Characterization and Preparation Facility (MCPF) at HKUST(GZ) provided necessary instruments for the crystal synthesis and characterizations. We acknowledge beamline I05 at Diamond Light Source for beamtime under proposal SI38254.

**Author contributions**

S.G., B.T.Z. and H.L. conceived the project. S.G. and H.L. led the ARPES measurement and analysis. R.G., X.W., Q.W., B.W., P. H. S.C., Z.S., T.K., C.C. and D.D. helped with the ARPES measurement. X.W., R.G., S.G. and H.L. performed the single crystal synthesis and the crystal characterizations. S.G., H.L. and B.T.Z. carried out the DFT and tight-binding model calculations. D.D., R.G., X.W., and C.Q. helped to give suggestions. S.G., B.T.Z. and H.L. did the majority of the paper writing, with contributions from all coauthors. B.T.Z. and H.L. directed the overall project.

**Competing interests**

The authors declare no competing interests.